\documentstyle{article}
\newcommand{\nid}{\noindent}

\newcommand{\be}{\begin{equation}} 
\newcommand{\ee}{\end{equation}} 
\newcommand{\ds}{\displaystyle}

\begin{document}

\title{{\small XIII Convegno Nazionale di Relativit\`a Generale e
Fisica della Gravitazione}\\
{\small 21-25 September 1998, Monopoli (Bari)}\\ 
{\bf Analytic solution of the Regge-Wheeler 
differential equation for black hole perturbations  
in radial coordinate and time domains}}

\author{{Alessandro D.A.M. Spallicci}\\
{Universit\`a del Sannio di Benevento}}

\date{}
\maketitle

{\abstract 
\nid{An analytic solution
of the  
Regge-Wheeler (RW) equation has been found via the Frobenius method   
at the regular singularity of the horizon $2M$, 
in the form of a time and radial coordinate dependent series.} 
The RW partial differential equation, 
derived from the Einstein field equations, 
represents the first order perturbations of the Schwarzschild metric. 
The known solutions are numerical in time domain or approximate and asymptotic  
for low or high frequencies in Fourier domain. 
The former is of 
scarce relevance for comprehension of the geodesic
equations for a body in the black hole field,
while the latter is  
mainly useful for the description of the emitted gravitational 
radiation. Instead 
a time domain solution is essential for the
determination of 
radiation reaction of the 
falling particle into the black hole, 
i.e. the influence of the emitted radiation on the motion  
of the perturbing mass in the black hole field. To this end, a semi-analytic 
solution of the inhomogeneous RW equation with the source term 
(Regge-Wheeler-Zerilli equation) shall be the next development.}
\vskip 5pt
\nid{\it Keywords: Partial differential equations (35Q75) Equations of
motion (83C10) Black holes (83C57)} 

\section{Introduction}

Regge and Wheleer \cite{RW1957} proved the vacuum stability of 
the Schwarzschild black hole, while  
Zerilli \cite{Z1970,Z1975}
studied the  
emitted radiation via the radial wave equation 
for polar perturbations, which source   
is a freely 
falling test mass $m$ towards the black hole of large 
mass $M$. 
Moncrief \cite{M1974} showed the gauge invariant significance of 
the wave equations.\\

Past work was concerned on the Fourier analysis 
of the emitted radiation (for a 
review see Ruffini \cite{R1972}) and not on the 
motion of the particle. Along the lines of radiation 
emission analysis, more recent work 
has been mainly developed in Japan (Osaka and Tokio), United States 
(Salt Lake City and   
Caltech), while symbolic computing on perturbations has been pursued in 
Canada (Kingston) and Italy (Roma, La Sapienza, ICRA).\\

Let us examine the known approaches to solutions of the RW equations: 
a Fourier antitrasform from an approximate 
frequency solution (where low or high frequencies are neglected, 
and thus also  
the corresponding antitransformed and yet unspecified 
time-parts of the  
trajectory of the incoming test mass) presents       
controversial issues of interpretation; a numerical time solution is 
not adequate for comprehension of the different contributions in the geodesic 
equations. \\
Therefore, an analytic solution, in time and obviously radial 
coordinate domains, remains the only way for
studying the 
problem of motion.\\

Past work was confined to a particle falling in   
an {\it unperturbed} Schwarz\-schild metric, but 
radiation reaction requires a geodesic path through   
the emission of radiation, in a {\it perturbed} Schwarzschild metric. \\
This novel approach for calculation of the motion and  
radiation reaction for 
the two-body problem (body plus particle, 
the small parameter $m/M$ being the ratio of the masses) is 
based on the following concept: 
in the background curvature given by the Schwarzschild 
geometry rippled by gravitational waves, the geodesic equations 
insure the presence of radiation reaction also 
for high velocities and strong field (the expansion parameter being $m/M$ 
and not $v/c$ or $\phi/c$).
This concept is    
applicable to any orbit, but radial fall is of 
interest due to the non-adiabatic regime (equality of 
radiation reaction and fall time scales), in which {\it the particle 
has to locally and immediately react to the emitted radiation}. 
The energy balance hypothesis 
(emitted radiation equal to the variation in the kinetic 
energy) 
is only used 
for the determination of the 4-velocity via the Lagrangian and 
for the normalization of divergencies;  
the solution in time domain of 
the Regge-Wheeler-Zerilli-Moncrief radial wave equation 
determines the metric tensor expressing the polar perturbations, in terms  
of which the geodesic equations are written; 
finally radiation reaction shall be identified
by subtraction of the 0th-order terms in the geodesic equations 
\cite{S1999a,S1999b}. \\
The long-term aim of this work is the identification of  
radiation reaction 
in the 
trajectory up to the horizon,  
without the assumption of adiabacity and 
with minimal use of the energy balance 
postulate.  \\
Future developments
may include: i) a general method for the determination of the 
motion of small objects 
in any orbit; ii) a contribution to the solution of the inhomogeneous   
2nd-order 
equations, which energy-momentum tensor is based on the geodesic equations 
{\it with perturbation effects} and yet to be found;  
iii) a post-Schwarzschildian formalism.\\
In this paper we are concerned with the partial fulfillment of the first
step, namely the solution of the associated homogeneous equation. \\

Radiation reaction is a fundamental concept 
in bodies motion theory, but also has relevant 
implications on detector's templates since the capture of 
stars by black holes is   
a source of gravitational waves.\\

For a brief critical review of the status of comprehension of radiation  
reaction and references, please refer to \cite{S1999a,S1999b}. 

\section{The equation}

The RWZM equation for polar perturbations is: 

\be
\frac {d^{2} \Psi_l(r,t)}{dr^{*2}} -  
\frac {d^{2} \Psi_l(r,t)}{dt^2} - V_{l}(r)\Psi_l (r,t) = S_{l}(r,t)
\label{eq:rwzm*}
\ee

where  

\be
r^{*}= r + 2M \ln \left ({\ds{r\over 2M}} - 1 \right )
\label{eq:rstarr}
\ee

 is the 
tortoise~coordinate and the potential 
$V_{l}(r)$ is:

\be
V_{l}(r) =  \left ( 1- \frac{2M}{r}\right ) 
\frac {2\lambda^{2}(\lambda + 1)r^{3} + 6\lambda^{2}Mr^{2} + 
18\lambda M^{2}r + 18M^{3}} 
{r^{3}(\lambda r + 3M)^2}
\ee

Further $\lambda = {1\over 2} (l - 1) (l + 2)$ and 
$S_{l}(r,t)$ is the $2^{l}$-pole 
source component and, for a radially 
falling particle, is expressed by:

{\small
\be
S_{l} =  
\frac 
{
\left(1-
{\ds\frac{2M}{r}}
\right ) k
}
{
(\lambda+1)(\lambda r+3M)
}
\left \{r
\left(1-
{\ds\frac{2M}{r}}
\right )^2 
\delta '[r- r_0(t)] 
-
\left[
(\lambda +1)-\frac{M}{r} 
- 
\frac{6Mr}{\lambda r+3M}
\right ]
\delta [r-r_0 (t)]
\right \}
\ee
}
where $k = 4m\sqrt{(2l + 1)\pi}$ 
and $r_0(t)$ 
(known geodesic equation of motion in {\it unperturbed Schwarzschild metric})
is the inverse of: 

\be
t = - 4M\left({\frac{r}{2M}}\right )^{1/2} 
- \frac{4M}{3}\left(\frac{r}{2M}\right )^{3/2} - 2 M
\ln \left [\left ( \sqrt{\frac{r}{2M}} - 1 \right ) 
\left (
\sqrt{\frac{r}{2M}} + 1 \right )^{-1} \right ] 
\label{eq:tofr} 
\ee

Eq. \ref{eq:tofr} reveals that the time domain spans from $-\infty$ to 
$+\infty$. 
The total energy radiated after a given $t_0$ per $l$ mode is:

\be
E = \frac{1}{64\pi}\frac{(l + 2)!}{(l-2)!}\int_{t_0}^\infty (\dot \Psi)^2dt
\ee
\section{The analytic solution}

Eq.(\ref{eq:rwzm*}) can be rewritten in terms of $t$ and $r^*$ (tortoise 
coordinate)
only, resulting into a partial differential equation    
with constant coefficients but solely    
via an approximate 
inverse function $r(r^*)$ of eq. (\ref{eq:rstarr}) and thus resulting
into   
an approximate p.d.e. Further, the solution is  
most interesting at $r^* = - \infty$ where a detailed analytic study is not  
possible. In  
the $(t,r)$ domain instead,
eq.(\ref{eq:rwzm*}) becomes:

\be
\frac{1}{A^2(r)}\frac {d^{2} \Psi_l}{dr^2} -  
\frac{1 - A(r)}{rA^2(r)}\frac {d \Psi_l}{dr} -  
\frac {d^{2} \Psi_l}{dt^2} - V_{l}(r)\Psi_l = S_{l}(r,t)
\label{eq:rwzm}
\ee

where
$A(r) = {\ds\frac{dr^*}{dr}} = {\ds\frac{r}{r - 2 M}}$.
The initial value problem is well defined 
($\Psi$ and $\dot\Psi$ are zero, i.e. 
particle at rest at infinity as the boundary conditions which affirm that 
the same functions are zero at infinity at all times). 
The associated homogeneous equation of eq. (\ref{eq:rwzm}) can be reduced to 
an 
ordinary differential equation with the position $\Psi (r,t) = R(r)T(t)$.    
Dividing eq. (\ref{eq:rwzm}) for $RT$ it is obtained that 
(dropping the l notation):  

\be
\frac{1}{RA^2}\frac {d^{2}R}{dr^2} -  
\frac{1 - A}{rRA^2}\frac {dR}{dr} -  
\frac{1}{T}\frac {d^{2} T}{dt^2} - V = 0
\label{eq:rwh}
\ee

The first, second and fourth term are not dependent upon $t$ and form  
time constants. Thus also the third term must be a constant in time and  
eq. (\ref{eq:rwh}) is autonomous \cite{Z}: 

\be
\frac{1}{T}\frac {d^{2} T}{dt^2} = \tau^2
\label{eq:rwht}
\ee

Let us pose $dT/dt = u$ and  
hence $d^2T/dt = uu_T$ where the lower subscript indicates a derivative 
to $T$. Eq. (\ref{eq:rwht}) becomes: 

\be
\frac{1}{T}uu_T = \tau^2
\ee

which solution is $ T = e^{\pm \tau}$. 
Eq. (\ref{eq:rwh}) becomes:

\be
\frac{1}{RA^2}\frac {d^{2}R}{dr^2} -  
\frac{1 - A}{rRA^2}\frac {dR}{dr} -  
\tau^2 - V = 0
\label{eq:rwh1}
\ee

or showing the dependence of $A(r)$ with $r$: 

\be
\frac {d^{2}R}{dr^2} +  
\frac{2M}{r(r-2M)}\frac {dR}{dr} -  
\frac{r^2R}{(r-2M)^2}(\tau^2 + V) = 0
\label{eq:rwh2}
\ee

For searching the solution, it is better to recast eq.
(\ref{eq:rwh2}) as:

\be
\frac {d^{2}R}{dr^2} +  
\frac{2M}{r(r-2M)}\frac {dR}{dr}  
- f(r)R = 0
\label{eq:rwh3}
\ee

where $f(r)$ is given by: 

{\small
\be
\frac{
\lambda^2 \tau^2 r^6
\!\!+\!\! 6 \lambda \tau^2 M r^5
\!\!+\!\! (9 \tau^2 M^2\!\!+\!\! 2 \lambda^3\!\!+\!\!2 \lambda^2) r^4
\!\!+\!\! 2 \lambda^2 M (1\!\!-\!\!2 \lambda) r^3
\!\!+\!\! 6 \lambda M^2 (3\!\!-\!\!2 \lambda) r^2
\!\!+\!\! 18 M^3 (1\!\!-\!\!2 \lambda) r
\!\!-\!\! 36 M^4
}
{
r^2(r-2M)^2(\lambda r + 3M)^2  
}
\ee
}

With the position $\rho = r - 2M$, eq. (\ref{eq:rwh3}) 
is finally recast as:

{\small
\be
(\rho + 2m)^2 [\lambda(\rho + 2M) + 3M]^2 
\frac {d^{2}R(\rho)}{d\rho^2} +  
\frac{2M(\rho + 2m) [\lambda(\rho + 2M) + 3M]^2}{\rho}
\frac {dR(\rho)}{d\rho} + \frac{\gamma(\rho)}{\rho^2}R(\rho) = 0
\label{eq:rwh4}
\ee
}
where 

\[
\gamma(\rho) = - [
\lambda^2 \tau^2 (\rho + 2M)^6
+ 6 \lambda \tau^2 M (\rho + 2M)^5
+ (9 \tau^2 M^2 + 2 \lambda^3 + 2 \lambda^2) (\rho + 2M)^4
\]
\be
+ 2 \lambda^2 M (1 - 2 \lambda) (\rho + 2M)^3
+ 6 \lambda M^2 (3 - 2 \lambda) (\rho + 2M)^2
+ 18 M^3 (1 - 2 \lambda) (\rho + 2M)
- 36 M^4 ]
\ee

Eq. (\ref{eq:rwh4}) may be solved with the Frobenius method \cite{vv}. 
Let us rewrite eq. (\ref{eq:rwh4}) as:

\be
\alpha(\rho)
\frac {d^{2}R(\rho)}{d\rho^2}  
+ \frac{\beta (\rho)}{\rho} \frac {dR(\rho)}{d\rho} + 
\frac{\gamma(\rho)}{\rho^2}R = 0
\label{eq:rwh5}
\ee

where the expressions of $\alpha(\rho)$ and $\beta(\rho)$ are 
straightforwardly 
derived from (\ref{eq:rwh4}). The functions $\alpha$, $\beta$ and $\gamma$ 
are regular about the expansion point $\rho = 0$ and have the form: 

\be
\alpha(\rho) = \sum_{n=0}^{4}\alpha_n\rho^n
~~~~~~~
\beta(\rho) = \sum_{n=0}^{3}\beta_n\rho^n
~~~~~~~
\gamma(\rho) = \sum_{n=0}^{6}\gamma_n\rho^n
\ee

The indicial equation ($\alpha_0 s^2 + (\beta_0 - \alpha_0)s + \gamma_0 = 0$) 
is: 

\be
4M^4 (2\lambda + 3)^2 s^2 + 
[4M^4 (2\lambda + 3)^2 - 4M^4(2\lambda + 3)^2)s  
- 16K^2M^6 (2\lambda + 3)^2 
= 0
\ee

and reveals that the roots are $s_{1,2} = \pm 2KM$ and differ by an integer  
($4KM$). One solution $R_a (\rho)$ is a series of the 
form: 

 \be
R_a(\rho) = 
\rho^s \sum_{n=0}^{\infty} R_0 \rho^n
\ee

while the other is of the form: 

{\small
\be
R_b(\rho) = 
\ln \rho \left [\sum_{n=0}^{\infty}(s+2KM)R_n(s)\rho^{n+s}\right ]_{s = -2KM}
+ \rho^{-2KM}\sum_{n=0}^{\infty}\frac{d}{ds}
\left [(s+2km)R_n(s)\right ]_{s = - 2KM} \rho^n
\ee
}

$R_0$ has an arbitrary value and all other terms depend upon it, e.g.:   

\be
R_1 = -4 \frac{M^3[(-16\lambda^2 - 102\lambda - 36)K^2M^2
                  +(8\lambda^2 +12\lambda +9)KM 
                  +(4\lambda^2 +10\lambda +9)]}
                  {4KM + 1}R_0
\ee

\section{Acknowledgements}

Thanks are due to NIKHEF (Nationaal Instituut voor Kernfysica en Hoge
Energie Fysika) and FOM (stichting voor Fundamenteel Onderzoek der
Materie) for their 
Research Grant in Mathematical Physics at the Theory section of
NIKHEF-Amsterdam; to Prof. I.W. Roxburgh of the Astronomy Unit, School
of
Mathematical Sciences, Queen Mary \& Westfield College, London; finally the      
the access to the computing facilities 
of the Mathematics Dept. "G. Castelnuovo" of Roma I 
University, La Sapienza is acknowledged.

\end{document}